\documentclass[usenatbib]{mnras}
\usepackage{txfonts}
\usepackage[utf8x]{inputenc}
\usepackage{graphicx}
\usepackage{natbib}
\usepackage{url}
\usepackage{amssymb}
\title[High Gas surface densities]{High Gas Surface Densities yet Low UV Attenuation in $z\sim1$ Disc Galaxies}

\author[R. Nordon]{
Raanan Nordon,$^{1}$\thanks{nordon@astro.tau.ac.il}
\\
$^{1}$School of Physics and Astronomy, Faculty of Exact Sciences, Tel-Aviv University, Tel-Aviv 69978, Israel. \\
}

\begin{document}
\maketitle

\begin{abstract}
The gas in galaxies is both the fuel for star formation and a medium that attenuates the light of the young stars.
We study the relations between UV attenuation, spectral slope, star formation rates, and molecular gas surface densities in a sample of 28 $z\sim1$ and a reference sample of 32 $z\sim0$ galaxies that are detected in CO, far-infrared, and rest frame UV.
The samples are dominated by disc-like galaxies close to the main SFR--mass relation.
We find that the location of the $z\sim1$ galaxies on the IRX-$\beta$ plane is correlated with their gas-depletion time-scale $\tau_{\rm dep}$ and can predict $\tau_{\rm dep}$ with a standard deviation of 0.16~dex.
We use IRX-$\beta$ to estimate the mean total gas column densities at the locations of star formation in the galaxies, and compare them to the mean molecular gas surface densities as measured from CO.
We confirm previous results regarding high $N_{\rm H}/A_{\rm V}$ in $z\sim1$ galaxies.
We estimate an increase of the gas filling factor by a factor of 4--6 from $z\sim0$ to $z\sim1$ and a corresponding increase of factor 3--2 in the mean column densities of the star forming clouds.
After accounting for the filling factor, the $z\sim1$ and the $z\sim0$ samples exhibit similar attenuation properties.
These indicate to similar porous geometries to the molecular clouds in star-forming disc galaxies at $0<z\lesssim1$.
\end{abstract}

\begin{keywords}
Galaxies:evolution -- Galaxies:ISM -- Galaxies:star formation -- Cosmology:observations -- Infrared:galaxies -- Ultraviolet:galaxies
\end{keywords}

%%%%%%%%%%%%%%%%%%%%%%%%%%%%%%%%%%%%%%%%%%%%%%%%
%
\section{Introduction} \label{sec:Introduction}
%
%%%%%%%%%%%%%%%%%%%%%%%%%%%%%%%%%%%%%%%%%%%%%%%%

In the context of star formation in galaxies, gas is usually discussed either as the medium that obscures the light emitted from the young stars, or as the fuel material for forming stars.
Attenuation towards a point source is a function of the gas column density \citep[e.g.,][]{Bohlin1978}.
Star formation rate (SFR) surface density is known to be correlated with the gas surface density \citep[e.g.,][]{Kennicutt89, Bigiel08, Tacconi10}.
In the usual terminology, the distinction between column density and gas surface density is a matter of spatial averaging, the later usually refers to larger spatial scales.
However, when we refer to the attenuation of the integrated emission from numerous point sources (young stars) in a galaxy, the column densities associated with the attenuation are spatial averages and the distinction gets blurred.
Therefore, one may expect that attenuation, the gas surface density, and the SFR in galaxies should correlate with each other, at least over spatial scales that include numerous star forming regions and molecular clouds.

Gas and dust in the interstellar medium (ISM) attenuate the stellar light as it passes through it.
The magnitude of this attenuation is wavelength dependent and thus, in addition to the extinction of stellar light, the spectral slope also changes.
The ultra-violet (UV) spectrum of a stellar population long-ward of the Lyman edge is usually approximated as a power law $f_{\lambda} \propto \lambda^\beta$, where $\beta$ is the spectral slope.
The UV continuum, classically measured between the FUV band (center wavelength $\sim$1600~\AA) and the NUV band (center wavelength $\sim$2800~\AA) of GALEX is dominated by the emission from young stars and serves as an important SFR indicator, especially at high redshifts where other tracers are unavailable, while the UV is conveniently redshifted into the optical bands.

The relation between the gas column density, spectral slope (or color excess in the visible wavelengths), and the attenuation has been the subject of study for many decades.
With the advent of far-infrared (FIR) telescopes, the infrared excess (IRX) defined as the ratio of the FIR to UV luminosities $L_{FIR}/L_{\rm UV}$ could be measured. IRX provides a direct measurement of the UV attenuation, under the assumption that the absorbed UV energy is re-emitted by the dust in the FIR.
Since the works of \citet{Meurer99} and \citet{Calzetti00} on the IRX--$\beta$ relation, a large body of work has been written on the subject, for example:
\citet{Charlot00}, \citet{Kong04}, \citet{Buat05}, \citet{Seibert05}, \citet{Howell10}, \citet{Hao11}, \citet{Wild11}, \citet{Overzier11}, \citet{Nordon2013}, \citet{Casey2014}, \citet{Alvarez2016}, and others.
These works show that the IRX--$\beta$ relation is not universal and that different kinds of galaxies follow different relations and with significant scatter in most samples.

Stars form in dense cores within molecular clouds.
While the clouds properties and number of young stars residing in each may vary quite a bit, on galactic scales a correlation has been found between the surface density of the molecular gas $\Sigma_{\rm H2}$ and the star formation rate surface density $\Sigma_{\rm SFR}$.
This relations is often referred to as the `Kennicutt-Schmidt' (KS) relation \citep{Kennicutt89, Schmidt59}.
The ratio of the two surface densities is the molecular gas depletion time-scale ($\tau_{\rm dep}=\Sigma_{\rm H2}/\Sigma_{\rm SFR}$) or its inverse, the star formation efficiency (SFE=$\tau_{\rm dep}^{-1}$).
In galaxies that follow the SFR-mass relation \citep{Brinchmann04, Noeske07, Elbaz07, Daddi07a} also referred to as the `main sequence' of star-forming galaxies, 
the gas depletion time-scale is typically of the order of $\tau_{\rm dep} \approx 1$~Gyr \citep{Leroy08, Wilson09, Daddi10, Tacconi13} with a weak dependence on redshift \citep{Genzel2015}, though \citet{Scoville2016} suggest a steeper dependency.
If the gas that correlates with the star formation is also the gas responsible for attenuating the UV from the newly formed stars, then one may naively expect that attenuation will correlate with gas surface density.
However, geometry, i.e. the arrangement of the stars and gas within the galaxy volume also plays a role.

\citet[][N13 henceforth]{Nordon2013} studied the relation between $\tau_{\rm dep}$ and the galaxy location in the $A_{\rm IRX}$--$\beta$ diagram, where $A_{\rm IRX}$ is the UV attenuation derived from the IRX value, as defined in N13 (see also \S~\ref{sec:AIRX-beta} below).
The scatter in the $A_{\rm IRX}$--$\beta$ is interpreted in N13 as a result of integrating the UV from numerous sources at varying optical depths in each galaxy, and the particular $A_{\rm IRX}$--$\beta$ location depends on the `attenuation distribution' $f(A)$ of the sources in each galaxy.
It is also possible to interpret  the location on this diagram as dependent on the extinction-law, that is in turn a result of the dust grains size distribution.
In N13 and in this work we deal with massive galaxies of near solar metallicity, that follow the main sequence of star forming galaxies.
Thus, we will assume that geometry has the larger effect and that the extinction-law varies little between the galaxies.

Previous works have found some inconsistencies when attempting to reconcile the SFR, the gas content, and the attenuation in $z>1$ galaxies.
\citet{Wuyts11b} attempted an exercise on a very large sample of galaxies in which they used the KS relation to convert the measured SFRs into gas surface densities and from that to predict the IRX values in the sample.
While this exercise was successful in galaxies of redshifts $z\lesssim1$, higher redshift galaxies had significantly lower observed IRX (lower UV attenuation) than that expected from their SFR surface densities.
\citet{Genzel2013} investigated a spatially resolved maps of a $z\sim1.5$ galaxy in CO, H$\alpha$, and {\it Hubble} photometry.
They found that the distribution (at $\sim$2~kpc resolution) of H$\alpha$ optical depths match a geometry in which the stars and gas are mixed together.
However, the value of $N_{\rm H}/A_{\rm V}$ required for the fit was nearly 5 times higher than the Milky-Way value, which is equivalent to the \citet{Wuyts11b} result.

In this work we will use the largest available sample of galaxies observed in CO, FIR, and optical bands in order to study the relation between $A_{\rm IRX}$--$\beta$, SFR, and molecular gas mass.
We will attempt to verify and generalize the results of \citet{Genzel2013} and \citet{Wuyts11b}, regarding the seemingly low $N_{\rm H}/A_{\rm V}$ at $z\sim1$.
In \S~\ref{sec:Data_and_Samples} we describe the samples used and their selection.
In \S~\ref{sec:Method} we verify the results of N13 using the new data.
In \S~\ref{sec:Columns_and_Densities} we use the $A_{\rm IRX}$--$\beta$ diagram to model and derive the gas surface densities and compare them to the directly observed CO columns.
In \S~\ref{sec:discussion} we discuss the results and our interpretation of them.

This work broadly follows the terminology and methods described in the N13 paper.
Throughout this paper we assume a \citet{Chabrier03} initial mass function (IMF), and a cosmology with ($\Omega_m$,$\Omega_\Lambda$,$H_0) = (0.3,0.7,70$~km~s$^{-1}$~Mpc$^{-1}$).

%%%%%%%%%%%%%%%%%%%%%%%%%%%%%%%%%%%%%%
%
\section{Data and Samples} \label{sec:Data_and_Samples}
%
%%%%%%%%%%%%%%%%%%%%%%%%%%%%%%%%%%%%%%

\begin{table*}
\caption{The $z\sim1$ sample - UV and FIR properties.}
\label{tab:z1_sample_1}
\begin{tabular}{ccccccccccc}

\hline

ID & CO sample$^{\dag}$ & z & $\beta$ & $\delta \beta$ & $L_{\nu}(1600)$ & $\delta L_{\nu}(1600)$ & $LIR^{\ddag}$ & $\delta LIR$ & $SFR^{\S}$ & $\delta SFR$ \\
 & & & & &$10^{28}\,erg s^{-1} Hz^{-1}$&$10^{28}\,erg s^{-1} Hz^{-1}$&$10^{12}\,L_{\odot}$&$10^{12}\,L_{\odot}$&$M_{\odot} yr^{-1}$&$M_{\odot} yr^{-1}$ \\

\hline

EGS12004280 & T13 & 1.023 &-0.24 &0.43 &2.07 &0.42 &0.77 &0.08 &86 &9 \\
EGS12007881 & T13 & 1.16 &-0.63 &0.16 &13.07 &0.79 &0.55 &0.08 &72 &9 \\
EGS12020405 & T13 & 1.379 &-0.90 &0.13 &6.59 &0.31 &1.29 &0.12 &147 &13 \\
EGS12024866 & T13 & 1.002 &-0.21 &0.13 &2.59 &0.16 &0.23 &0.06 &27 &7 \\
EGS12028325 & T13 & 1.159 & -1.19 &0.42 &10.37 &1.62 &0.60 &0.08 &74 &9 \\
EGS13003805 & T13 & 1.23 &0.59 &0.15 &1.18 &0.09 &1.30 &0.12 &143 &13 \\
EGS13004291 & T13 & 1.145 &-0.46 &0.15 &4.56 &0.31 &4.71 &0.10 &519 &11 \\
EGS13004661 & T13 & 1.192 &-0.24 &0.08 &3.43 &0.13 &0.50 &0.13 &58 &14 \\
EGS13011155 & T13 & 1.012 &-0.77 &0.12 &2.34 &0.13 &1.64 &0.05 &181 &6 \\
EGS13011166 & T13 & 1.529 &-0.68 &0.33 &10.21 &1.09 &2.65 &0.18 &299 &20 \\
EGS13011439 & T13 & 1.099 &-0.54 &0.34 &3.05 &0.46 &0.53 &0.10 &60 &11 \\
EGS13017614 & T13 & 1.18 &0.43 &0.09 &2.12 &0.08 &0.73 &0.12 &82 &13 \\
EGS13017707 & T13 & 1.037 &1.20 &0.39 &0.14 &0.03 &2.98 &0.06 &326 &6 \\
EGS13018312 & T13 & 1.105 &1.88 &0.54 &0.25 &0.07 &0.43 &0.11 &47 &12 \\
EGS13018632 & T13 & 1.229 &0.44 &0.36 &0.86 &0.16 &0.93 &0.08 &103 &8 \\
EGS13026117 & T13 & 1.241 &0.02 &0.10 &2.88 &0.14 &0.92 &0.10 &103 &11 \\

BzK-4171  & D10  & 1.465 &0.44 &0.11 &0.66  &0.04 &0.92 &0.05 &102 &6 \\
BzK-21000 & D10  & 1.523 &-0.60 &0.10 &2.42 &0.12 &2.10 &0.07 &231 &8 \\
BzK-16000 & D10  & 1.522 &0.06 &0.06 &2.74  &0.09 &0.73 &0.09 &83  &10 \\
BzK-17999 & D10  & 1.414 &-0.01 &0.12 &0.89 &0.05 &1.07 &0.02 &118 &2 \\
BzK-12591 & D10  & 1.6   &-0.13 &0.05 &4.00 &0.12 &2.41 &0.08 &267 &9 \\

PEPJ123712+621753 & M12 & 1.249 &-0.50 &0.12 &1.34 &0.09 &0.42 &0.03 &47 &4 \\
PEPJ123709+621507 & M12 & 1.224 &-0.84 &0.09 &3.59 &0.17 &0.31 &0.04 &37 &4 \\
PEPJ123759+621732 & M12 & 1.084 &-1.15 &0.17 &6.58 &0.64 &0.36 &0.04 &45 &5 \\
PEPJ123721+621346 & M12 & 1.021 &-1.19 &0.17 &1.83 &0.18 &0.54 &0.03 &60 &3 \\
PEPJ123633+621005 & M12 & 1.016 &-0.92 &0.17 &3.59 &0.35 &0.98 &0.03 &110 &3 \\
PEPJ123646+621141 & M12 & 1.016 &-1.01 &0.17 &6.97 &0.67 &0.41 &0.03 &51 &3 \\
PEPJ123750+621600 & M12 & 1.17  &0.63 &0.11 &1.86 &0.12 &0.26 &0.04 &30 &4 \\

\hline
\end{tabular}

\flushleft
$\dag$ T13 are from \citet{Tacconi13}, D10 are from \citet{Daddi10}, M12 are from \citet{Magnelli12}. \\
$\ddag$ the total luminosity between 8--1000 $\AA$.\\
$\S$ Total ${\rm SFR = SFR_{IR}+SFR_{UV}}$, see Eq.~\ref{eq:SFR_IR} \& Eq.~\ref{eq:SFR_UV}.\\

\end{table*}

\begin{table*}
\caption{The $z\sim1$ sample - Gas properties.}
\label{tab:z1_sample_2}
\begin{tabular}{ccccccccc}
\hline

ID & $\log M_{mol}(IRX)$ & ${\delta}\log\,M_{mol}(IRX)^{\dag}$ & $\log\,M_{mol}(CO)^{\ddag}$ & $\delta\log\,M_{mol}(CO)^{\ddag}$ & $\log\,\Sigma_{mol}^{\ddag}$ & $A_{min}$ & $\Delta A$ & $\log N_{H}/A_{V}$ \\
 & $M_{\odot}$ & dex & $M_{\odot}$ & dex & $M_{\odot} pc^{-2}$ & mag & mag & $cm^{-2}mag^{-1}$\\

\hline

EGS12004280 &10.75 &0.12 &10.85 &0.08 &2.7 &0.97 &21.08 &21.92 \\
EGS12007881 &10.69 &0.07 &10.88 &0.03 &2.57 &1.48 &1.12 &22.55 \\
EGS12020405 &10.87 &0.05 &11.04 &0.08 &2.83 &0.15 &24.07 &22.03 \\
EGS12024866 &10.36 &0.12 &10.51 &0.07 &2.38 &1.40 &3.43 &22.17 \\
EGS12028325 &10.56 &0.13 &10.20 &0.02 &2.37 &0.0&10.75 &21.92 \\
EGS13003805 &11.05 &0.06 &11.32 &0.01 &3.03 &2.02 &23.46 &22.18 \\
EGS13004291 &11.36 &0.04 &11.54 &0.01 &3.76 &0.70 &73.93 &22.47 \\
EGS13004661 &10.65 &0.12 &10.52 &0.01 &2.32 &1.02 &8.14 &21.90 \\
EGS13011155 &10.87 &0.03 &11.04 &0.08 &2.48 &0.32 &71.57 &21.21 \\
EGS13011166 &11.21 &0.10 &11.41 &0.03 &2.99 &0.43 &24.43 &22.17 \\
EGS13011439 &10.59 &0.11 &10.76 &0.08 &3.13 &0.61 &14.01 &22.53 \\
EGS13017614 &10.90 &0.08 &11.04 &0.05 &2.94 &1.85 &8.68 &22.43 \\
EGS13017707 &11.13 &0.11 &10.97 &0.01 &3.06 &2.79 &213.93 &21.30 \\
EGS13018312 &10.84 &0.18 &10.81 &0.01 &2.81 &3.69 &7.81 &22.21 \\
EGS13018632 &10.87 &0.10 &10.53 &0.01 &3.18 &1.84 &27.49 &22.27 \\
EGS13026117 &10.91 &0.06 &11.15 &0.01 &3.34 &1.31 &13.26 &22.72 \\

BzK-4171  & 10.84 &0.04 &10.97 &0.05 &3.03 &1.84 &35.06 &22.03 \\
BzK-21000 & 10.99 &0.03 &10.99 &0.05 &2.61 &0.53 &73.20 &21.32 \\
BzK-16000 & 10.84 &0.06 &10.85 &0.05 &2.64 &1.38 &10.54 &22.10 \\
BzK-17999 & 10.80 &0.03 &10.88 &0.05 &2.74 &1.27 &51.33 &21.59 \\
BzK-12591 & 11.21 &0.02 &11.16 &0.08 &3.06 &1.11 &29.70 &22.14 \\

PEPJ123712+621753 & 10.44 &0.05 &10.21 &0.11 &2.92 &0.65 &23.95 &22.10 \\
PEPJ123709+621507 & 10.34 &0.05 &10.47 &0.09 &2.74 &0.24 &10.21 &22.29 \\
PEPJ123759+621732 & 10.36 &0.06 &10.56 &0.11 &1.69 &0.15 &7.39  &21.39 \\
PEPJ123721+621346 & 10.38 &0.05 &10.46 &0.15 &2.60 &0.09 &37.64 &21.61 \\
PEPJ123633+621005 & 10.71 &0.04 &10.76 &0.11 &2.12 &0.13 &33.68 &21.17 \\
PEPJ123646+621141 & 10.45 &0.05 &10.46 &0.10 &1.65 &0.14 &8.03  &21.32 \\
PEPJ123750+621600 & 10.59 &0.07 &10.71 &0.06 &2.33 &3.52 &0.07  &22.07 \\

\hline
\end{tabular}

\flushleft
$\dag$ Formal error only, carried over from the photometry. \\
$\ddag$ From the corresponding CO-sample paper. See table~\ref{tab:z1_sample_1}. \\

\end{table*}

\subsection{Sample Selection}
\label{sec:Sample_Selection}

In order to study the relation between the effective UV attenuation, the UV color and the gas column, we require UV photometry, FIR photometry, and CO gas mass measurements.
N13 used a sample of $z>1$ sources with CO measurements that were also detected in the FIR by {\it Herschel}.
The galaxies in the N13 sample are disc-like galaxies in the optical images and lie close to the SFR--$M_{\rm star}$ relation at their redshifts.
Since these sources lie in the deep extragalactic fields GOODS-N and the Extended Groth Strip (EGS), they also have optical (rest frame UV) photometry available.

Briefly mentioning, the CO data for the N13 $z\sim1$ sample was compiled from \citet{Daddi10}, \citet{Tacconi10}, and \citet{Magnelli12}.
The {\it Herschel} FIR and {\it Spitzer} 24~$\mu$m photometry in GOODS-N is from the combined data of the PEP project \citep{Lutz11} and GOODS-Herschel \citep{Elbaz11}.
The details of the combined reduction are given in \citet{Magnelli13}.
{\it Herschel}-PACS and {\it Spitzer} 24~$\mu$m photometry in EGS is from the catalogues of the PEP project.
UBVIz photometry in GOODS-N is from the optical catalogue used by PEP which is a compilation of various optical GOODS-N catalogues.
All PEP project products are publicly available \footnote{\url{http://www.mpe.mpg.de/ir/Research/PEP/public_data_releases}}.
Optical photometry in EGS is from the publicly released catalogues of the AEGIS\footnote{\url{http://aegis.ucolick.org/astronomers.html}} team \citep{Davis2007}.

In this work we use the same CO and FIR detected $z\sim1$ sample from N13 and add the galaxies from the PHIBSS project \citep[][T13 hereafter]{Tacconi13} that have a counterpart in the {\it Herschel}-PACS 100 and 160 $\mu$m EGS images from the PEP project, and good rest-UV (1600--2800 $\AA$ region) photometry.
About half (20) of the PHIBSS galaxies are detected in FIR.
A requirement for at least two bands that observe rest frame UV (to allow a power-law fit), reduced our PHIBSS sub-sample size to 17.
We further eliminated EGS12004351 from our sample due to its very red colors, and its image that indicates an edge-on galaxy undergoing a merger or strong interaction (thumbnail images available in T13).
We are thus left with 16 PHIBSS galaxies in our sample.

Six of the PHIBSS galaxies \citep[originally from][]{Tacconi10} were already included in the N13 sample.
For these we use the CO measured values from the new reduction of T13.
The PHIBSS sample is dominated by disc-like galaxies that lie near the SFR--$M_{\rm Star}$ relation.
While the PHIBSS sample was not specifically selected this way, their requirements for the detection of certain emission lines favour the selection of face-on disc galaxies.
In total, we added 10 new PHIBSS sources to the N13 CO-detected $z\sim1$ sample.
The current $z\sim1$ sample has 28 sources as detailed in tables \ref{tab:z1_sample_1} \& \ref{tab:z1_sample_2}.

In order to compare the $z\sim1$ galaxies to local galaxies we use the COLD GASS \citep{Saintonge11a} sub-sample that was used in N13.
COLD GASS is a low redshift, unbiased, mass-selected sample observed with the IRAM 30~m telescope in CO(1-0).
Out of this sample N13 selected the galaxies with GALEX UV and IRAS FIR detections.
A small fraction of mergers is naturally included in this sub-sample (3/32) and the rest are representative of low-$z$ galaxies around the SFR-$M_{\rm star}$ relation.
We refer to this sample as the "$z\sim0$ sample".
%The details of the galaxies in this sample are given in {\bf Table @@}.

% % % % % % % % % % % % % % % % % % % % % % % % % % % % % % % % % % %
\subsection{Measuring $L_{\rm IR}$, $L_\nu(1600\AA)$, and $\beta$}
\label{sec:measuring_LIR_L1600_beta}
% % % % % % % % % % % % % % % % % % % % % % % % % % % % % % % % % % %

The total infrared luminosity in the 8--1000 $\mu$m range is calculated by fitting a \citet{CE01} template to the {\it Herschel} 160 $\mu$m, 100 $\mu$m, and {\it Spitzer} 24~$\mu$m photometry, allowing both spectral energy distribution (SED) template and scale to vary, as described in N13.
We only use templates characterized in the library as templates of $L_{\rm IR}>1\times10^{10} \, L_{\rm sun}$, though as said, we allow their normalization to vary.
The template that achieved the lowest $\chi^2$ is selected and the range of $L_{\rm IR}$ solutions from all templates that achieved $\chi^2 < \chi_{\rm min}^2+1$ determines the $1\sigma$ uncertainty $\delta L_{\rm IR}$.
By using a range of SED shapes typical of local spirals, luminous, and ultra-luminous galaxies (LIRGs \& ULIRGs) with free scaling (varying $L_{\rm IR}$), we allow optimal templates to be fitted to the $z>1$ galaxies, that tend to have a different association between template shape and $L_{\rm IR}$ than the local galaxies \citep{Nordon12}.

We derive $\nu L_\nu(1600\AA)$ luminosity at 1600~$\AA$ and the UV spectral slope $\beta$ by fitting a power-law SED. 
We select all available magnitude measurements for the source through filters whose central wavelength observes between 1300(1+$z$) and 3000(1+$z$) \AA.
Let $\beta$ be the spectral slope around 1600~\AA, such that $\nu L_{\nu}(\lambda) \propto \lambda^{\beta+1}$ at rest frame.
For each source, we pass a red-shifted power-law spectrum through the filter responses and minimize $\chi^2$ by adjusting $\beta$ and the luminosity at rest 1600~\AA.
The fit results for the UV and IR are given in table~\ref{tab:z1_sample_1}.

% % % % % % % % % % % % % % % % % % % % % % %
%
\subsection{Measuring $A_{\rm IRX}$}
\label{sec:AIRX-beta}
%
% % % % % % % % % % % % % % % % % % % % % % %
Traditionally, IRX is defined as the ratio of the FIR to FUV fluxes, where FIR is measured through a far infrared filter at 60~$\mu$m or longer wavelength, and FUV is measured with a filter covering the 1600~\AA\ region.
Here, we follow the definitions of N13 and thus the FIR luminosity $L_{\rm IR}$ refers to the total 8--1000 $\mu$m integrated luminosity and FUV is ${\nu}L_{\nu}$ at rest 1600~\AA.
Both the UV and the FIR luminosities are used as SFR indicators \citep{Kennicutt98}.
The two complete each other since the fraction of the UV luminosity from the young stars that is absorbed by the dust is re-emitted in the FIR by the dust.
Thus, it is commonly assumed that ${\rm SFR} = {\rm SFR_{UV}} + {\rm SFR_{IR}}$, where ${\rm SFR_{UV}}$ is derived from the observed UV luminosity without an attenuation correction.
The conversions from luminosities to SFR are from \citet{Kennicutt98}, and we divide the SFR by a factor of 1.6 to convert from a Salpeter to a \citet{Chabrier03} initial mass function (IMF)
\begin{equation}
 {\rm SFR}_{\rm IR} = 1.09\times10^{-10} \frac{L_{\rm IR}}{L_\odot}
 \label{eq:SFR_IR}
\end{equation}
\begin{equation}
 {\rm SFR}_{\rm UV} = 8.75\times10^{-29} \frac{L_\nu(1600\,{\rm \AA})}{\rm erg s^{-1} cm^{-2} Hz^{-1}} \, .
 \label{eq:SFR_UV}
\end{equation}

We define the effective UV attenuation $A_{\rm IRX}$ as the attenuation required in order to correct the star formation rate as estimated from the observed UV luminosity ${\rm SFR_{UV}}$ to the total SFR,
\begin{equation}
 A_{\rm IRX} = 2.5 \log\left( \frac{SFR_{\rm IR}}{SFR_{UV} } +1 \right) \quad.
 \label{eq:A_IRX_definition}
\end{equation}

%%%%%%%%%%%%%%%%%%%%%%%%%%%%%%%%%%%%%%
%
\section{IRX-$\beta$ and the gas depletion time-scale} \label{sec:Method}
%
%%%%%%%%%%%%%%%%%%%%%%%%%%%%%%%%%%%%%%

%%%%%%%%%%%%%%%%%%%%%%%%%%%%%%%%%
\subsection{The empirical relation}
%%%%%%%%%%%%%%%%%%%%%%%%%%%%%%%%%

Our photometric method for deriving molecular gas masses from FIR and UV has been presented in details in N13.
In this section we will briefly describe the main principles of the method.
The relation between $A_{\rm IRX}$ and the UV spectral slope $\beta$ depends on the geometrical arrangement of the young stars and the obscuring gas and dust.
The geometrical arrangement also affects the ratio between the number of young stars (proportional to the SFR) and amount of gas along the line of sight.
This leads to a typical relation between the molecular gas content, SFR and UV attenuation in normal star forming galaxies, empirically found to be (N13):
\begin{equation}
 \log\left( \frac{M_{\rm mol}}{10^{9}M_\odot} \right) = \log\left( \frac{A_{\rm IRX}{\rm SFR}}{M_\odot yr^{-1}} \right) -0.20(A_{\rm IRX}-1.26\beta) + 0.09
 \label{eq:M_mol_from_SA}
\end{equation}
An alternative way to present this relation is to move the log(SFR) term to the left hand side.
Thus, the molecular gas depletion time $\tau_{\rm dep}$ is given as a function of the location of the galaxy on the $A_{\rm IRX}$--$\beta$ diagram:
\begin{equation}
 \log\left( \frac{\tau_{\rm dep}}{\rm Gyr} \right) = \log(A_{\rm IRX}) -0.20(A_{\rm IRX}-1.26\beta) + 0.09
 \label{eq:tdep_from_SA}
\end{equation}
This means that at a constant $\beta$, when moving upwards in the diagram (increasing $A_{\rm IRX}$) we will tend to find galaxies with a shorter molecular-gas depletion time (i.e., a higher star formation efficiency).

N13 report that for galaxies with extreme sSFRs, such as local ULIRGs and $z>1$ sub-millimeter galaxies (SMGs) Eq.~\ref{eq:M_mol_from_SA} tends to over estimate the gas masses (and hence also $\tau_{\rm dep}$) by a significant factor and with a large scatter.
The original samples from which our current sample was pulled tend to avoid extreme cases of starbursts either by selection (in the $z\sim1$ samples), or by the natural rarity of such objects even though they are not selected against (in COLD~GASS).

%%%%%%%%%%%%%%%%%%%%%%%%%%%%%%%%%%%%%%%%%%%%%%%%%
\subsection{Testing against CO measurements}
\label{sec:test_CO_measurements}
%%%%%%%%%%%%%%%%%%%%%%%%%%%%%%%%%%%%%%%%%%%%%%%%%

In N13, Eq.~\ref{eq:M_mol_from_SA} has been calibrated against a sample of 18 $z\sim1$ galaxies.
It has also been found to correctly predict the molecular gas content in normal $z\sim0$ disc galaxies drawn from the COLD~GASS sample.
CO measurements are still the only way to directly observe molecular gas in distant galaxies and while such measurements may include some systematic uncertainties they are the best `yard stick' against which we test the accuracy of indirect methods.
The $z\sim1$ sample used in this work adds 10 PHIBSS galaxies that were not in N13 and 6 other N13 galaxies had their CO measurements updated by PHIBSS.
It is worthwhile to test the accuracy of Eq.~\ref{eq:M_mol_from_SA} \& \ref{eq:tdep_from_SA} on a sample that includes many galaxies that were not in the sample from which these relations were derived.

\begin{figure}
 \centering
 \includegraphics[width=\columnwidth]{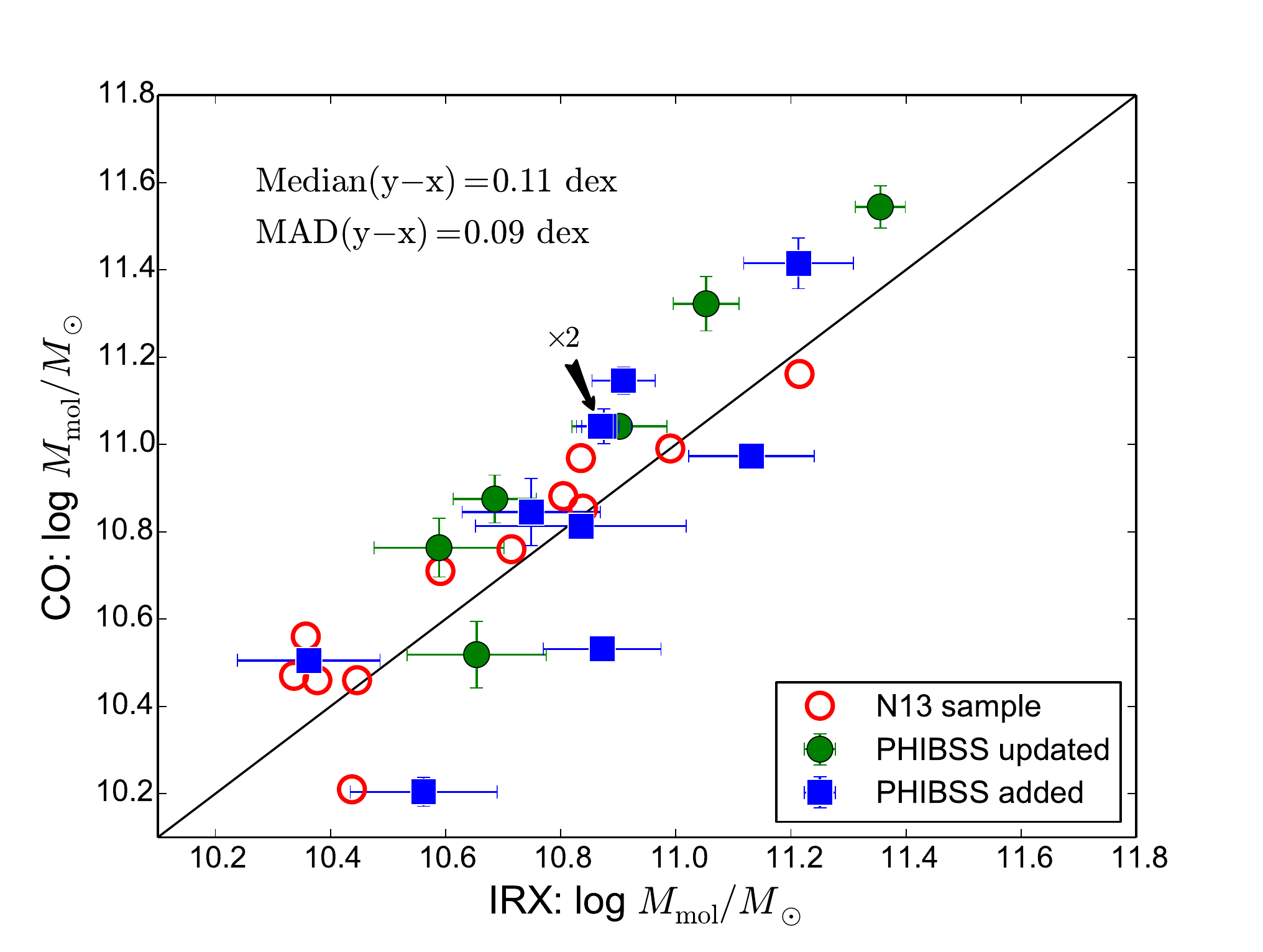}
 \caption{The molecular gas mass from CO measurements versus gas mass estimated with the IRX--$\beta$ method.
 N13 galaxies that are not included in the PHIBSS sample are marked with red empty circles.
 Six galaxies included in N13 that have CO measurements updated by PHIBSS are marked with green circles.
 Ten (two nearly on top of each other) new PHIBSS galaxies not included in N13 are plotted as blue squares. The median and median absolute deviation refer to the scatter around the 1:1 ratio.}
 \label{fg:Mmol_SA_vs_PHIBSS}
\end{figure}

In Figure~\ref{fg:Mmol_SA_vs_PHIBSS} we plot for the $z\sim1$ sample $M_{\rm mol}$ as measured from CO versus $M_{\rm mol}$ as estimated from UV and FIR photometry using Eq.~\ref{eq:M_mol_from_SA}.
The molecular gas masses are predicted by the IRX-$\beta$ empirical method, with a median bias of 0.11~dex (under prediction), and a median absolute deviation (MAD) of 0.09~dex (standard deviation, stdev of 0.16~dex) scatter.
This scatter is similar to the general accuracy of the method, estimated in N13 to have stdev of 0.16~dex.
When using the $A_{\rm IRX}$--$\beta$ to measure gas masses, this random error should in principle be added on top of the systematic uncertainty of the CO measurements that are used as calibrators.
Unfortunately, the latter is unknown.

\begin{figure}
 \centering
 \includegraphics[width=\columnwidth]{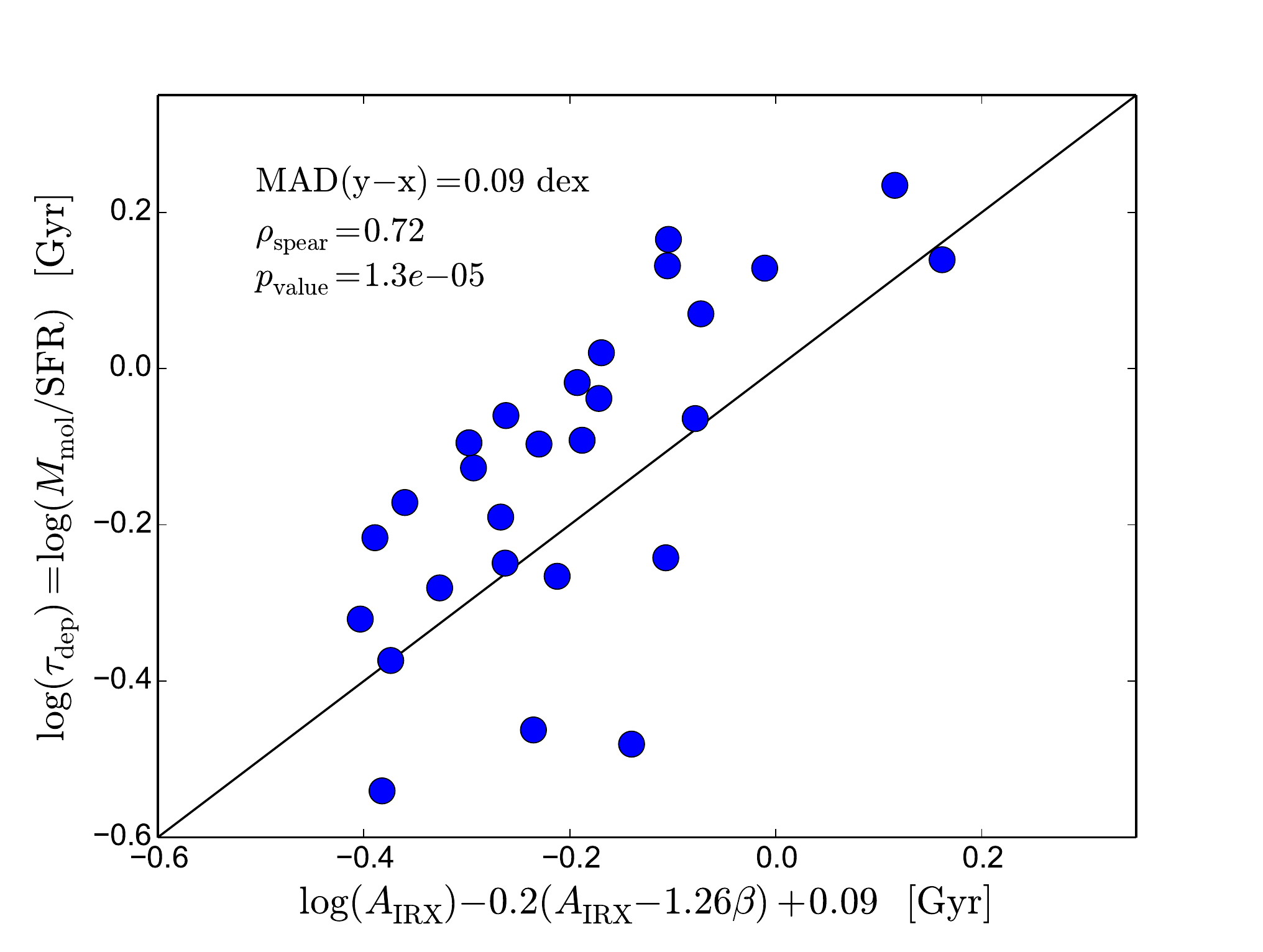}
 % tau_dep_vs_SA_estimate.eps: 0x0 pixel, 300dpi, 0.00x0.00 cm, bb=13 175 598 616
 \caption{
 $\tau_{\rm dep}$ as measured from CO and FIR versus $\tau_{\rm dep}$ as estimated from the location of the galaxies on the $A_{\rm IRX}$--$\beta$ diagram (Eq.~\ref{eq:tdep_from_SA}).
 }
 \label{fg:tau_dep_PHIBSS_vs_SA}
\end{figure}

A relevant question is whether the $A_{\rm IRX}$--$\beta$ information actually contributes anything to the accuracy of $M_{\rm mol}$, versus simply assuming a uniform $\tau_{\rm dep}$ and scaling the molecular gas mass according to the SFR (i.e., $M_{\rm mol} = {\rm SFR} \times \tau_{\rm dep}$).
The contribution of the IRX--$\beta$ information is to modify the assumed $\tau_{\rm dep}$ (Eq.~\ref{eq:tdep_from_SA}).
In Figure~\ref{fg:tau_dep_PHIBSS_vs_SA} we plot $\tau_{\rm dep}$ as measured from CO and SFR, versus the prediction from E.q.~\ref{eq:tdep_from_SA}.
As we can see from the figure, there is a relatively good correlation between the measured $\tau_{\rm dep}$ and the prediction from the $A_{\rm IRX}$ and $\beta$ values.
The Spearman ranking correlation coefficient is $\rho=0.72$ and the null hypothesis that the apparent correlation is a random result is ruled out with a probability of $p=1.3\times10^{-5}$.
The MAD is 0.09~dex (stdev 0.14~dex) over a dynamic range of 0.6~dex in $\tau_{\rm dep}$.
The measured $\tau_{\rm dep}$ on the other hand has an intrinsic scatter of 0.15~dex MAD (0.22~dex stdev).
This would also be the scatter in the derived gas masses had we assumed a uniform $\tau_{\rm dep}$ and multiplied by the SFR. 
That is 60\% larger scatter than the $A_{\rm IRX}$--$\beta$ prediction.

The above test demonstrates that a real connection exists between the obscuration of the young stars and the star formation efficiency in main-sequence galaxies.
Perhaps such a connection may seem trivial, however in the following sections we will attempt to look into this relation in more details and show that it may not be so.

%%%%%%%%%%%%%%%%%%%%%%%%%%%%%%%%%%%%%%
%
\section{Obscuring gas columns and mean surface densities} 
\label{sec:Columns_and_Densities}
%
%%%%%%%%%%%%%%%%%%%%%%%%%%%%%%%%%%%%%%

\subsection{The $A_{\rm IRX}$-$\beta$ plane and models grid}
\label{sec:IRX-beta_location}

The change in color or spectral slope of a source is traditionally associated with a foreground column of dust that reddens a point source, such as a star.
However, when integrating the light from a whole galaxy we observe a large collection of stars, each with its column of obscuring material.
The photometric magnitudes from which we determine the spectral slope are a weighted result of all these stars, where the least obscured ones will tend to contribute the most light and dominate the measured slope.

N13 describes this weighted contribution with the use of an `attenuation distribution' function $f(A)$, defined as the fraction of the UV emitting stars with an attenuation magnitude between $A$ and $A+dA$.
For example, a slab of gas with stars evenly distributed within it (classic `even-mix' geometry) will produce a flat $f(A)$ from zero up to $\Delta A$ the attenuation from one side of the slab to the other.
An additional gas and dust screen in front of the `even-mix' slab will result in a shift in $f(A)$ that will now span between $A_{\rm min}$ (the attenuation through the front screen) and $A_{\rm min}+\Delta A$.
One may also assume a symmetrical gas layer (without UV sources) on the far side of the `even-mix' slab.
Such `sandwich' models can also be found in the literature \citep[e.g.,][]{Wild11}.
For a more detailed discussion about $f(A)$ and the resulting location in $A_{\rm IRX}$--$\beta$ see N13.

We would now like to use the $A_{\rm IRX}$--$\beta$ data in order to estimate the typical gas surface density in the regions of star formation in the galaxies of our sample.
For this purpose we need the total UV attenuation through the gas disc from one side to the other, in these regions.
What we measure as $A_{\rm IRX}$ is an effective attenuation and its relation to the total attenuation through the galaxy disc depends on the geometry, or in other words on $f(A)$.

\citet{Genzel2013} analyzed spatially resolved images of a $z\sim1.5$ disc-like galaxy (EGS13011166, included in our sample) in CO, H$_\alpha$ and optical photometry.
They concluded that the local relations between the gas surface density and H$_\alpha$ optical depth agree with an `even mix' geometry.
We adopt this assumption for our sample, and also allow a possible foreground component.
In this classic `sandwich' model the young stars are mixed with gas and dust inside the galaxy disc, possibly in the form of star forming regions and giant molecular clouds (GMCs), and a diffuse component (the `bread slices'), possibly of atomic or ionized gas that does not form stars and envelopes the mixed component.
We do not however make any assumption regarding the nature of the obscuring gas in these components, except that H$_2$ dominates the total column density \citep{Sternberg2014, Nordon2016}.

\begin{figure}
 \centering
 \includegraphics[width=\columnwidth]{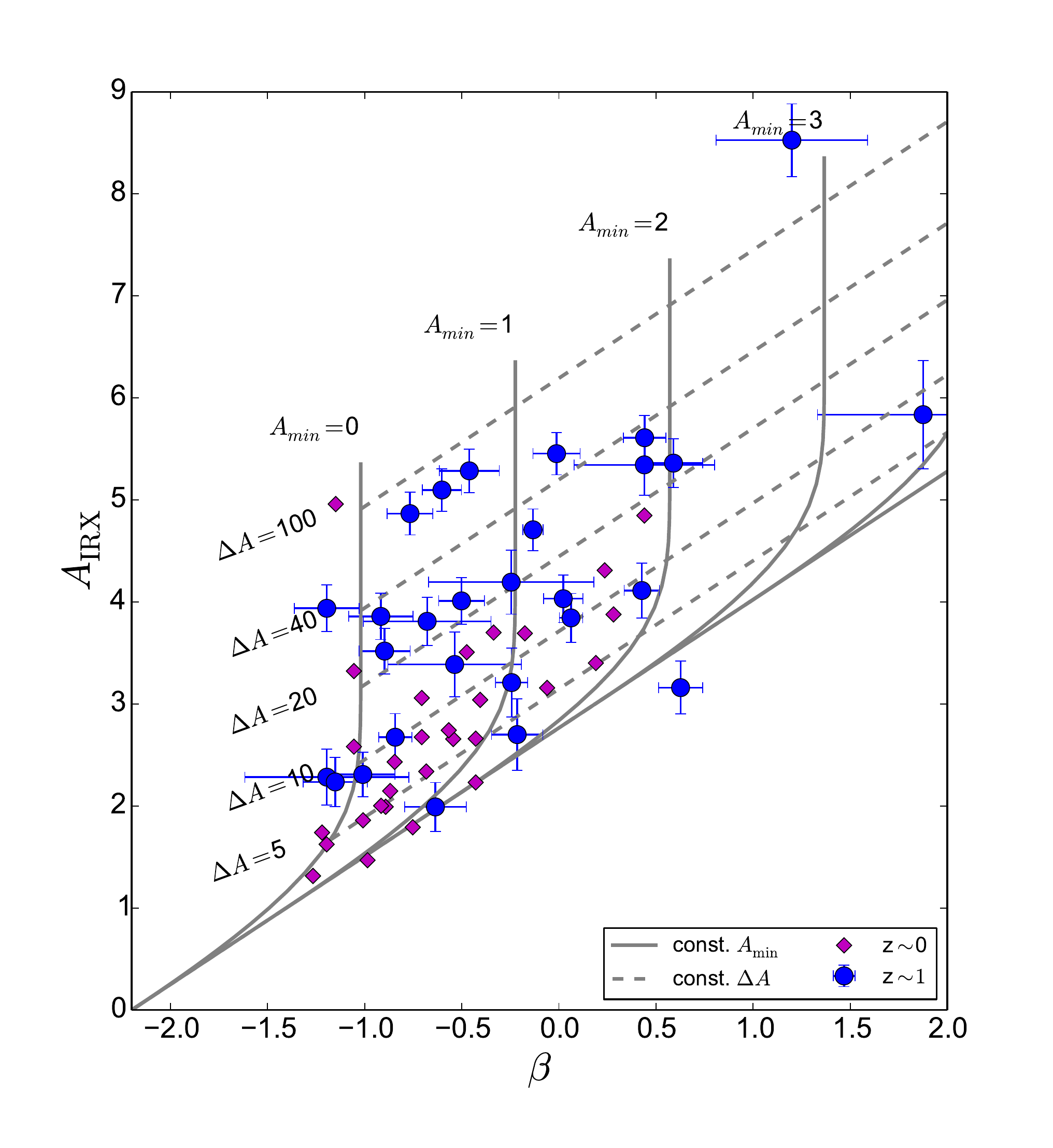}
 \caption{The $z\sim1$ sample in blue circles plotted on top of the `sandwich' even-mix plus foreground screen models grid.
 Solid gray lines indicate constant foreground attenuations ($A_{\rm min}$), while broken gray lines indicate constant even-mix slabs thickness ($\Delta A$).
 The $z\sim0$ sample is plotted in red diamonds for reference. }
 \label{fg:sandwich_grid}
\end{figure}

%Eq.~\ref{eq:AIRX_for_sandwich} \& Eq.~\ref{eq:beta_for_sandwich} above are the solutions for the specific case in which $f(A)$ resembles a flat distribution between $A_{\rm min}$ and $A_{\rm min}+\Delta A$.
The $A_{\rm IRX}$--$\beta$ solution for the case in which $f(A)$ resembles a flat distribution between $A_{\rm min}$ and $A_{\rm min}+\Delta A$ is given in N13 (their Eq.~13 \& Eq.~14).
In Figure~\ref{fg:sandwich_grid} we plot our $z\sim1$ sample on the $A_{\rm IRX}$-$\beta$ plane on top of a grid of constant $A_{\rm min}$ and constant $\Delta A$ models.
For comparison, we also plot the $z\sim0$ sample on this grid.
A few galaxies are just outside the area span by the models grid and in these cases we adopt the value of the nearest model in terms of standard deviations (less than 1.5~$\sigma$ shift in all cases).

We would like to point out a few features of the models grid.
The solid diagonal line that connects to the $\beta$ axis is the pure foreground screen model ($\Delta A=0$).
In that case $A_{\rm IRX}$ is simply the attenuation through the foreground screen ($A_{\rm min}$).
We assume that there is a similar screen on the other side of the stars, which does not affect the observed UV, but should be included in the total column of gas.
Increasing $A_{\rm min}$, i.e. a thicker foreground screen will move a modeled galaxy in parallel to this line.
The slope of this line is determined by the attenuation law $\gamma$ (i.e. $A_{\rm UV} \propto \lambda^\gamma$).
The line intersects the $\beta$ axis at $\beta_0$ the unobscured UV slope.
Varying $\beta_0$ shifts the models grid left/right accordingly.
The `even mix' component ($\Delta A$) contributes to increasing $\beta$, but quickly turns into a vertical line due to highly attenuated stars that contribute only to the IR.
For this reason a foreground screen component is needed in order to create galaxies with $\beta>-1$.
For $\Delta A >>1$, the effective attenuation increase with the log of the slab thickness, and $A_{\rm IRX} \approx A_{\rm min} + 2.5\log_{10}(0.921\Delta A)$.

As we can see in Figure~\ref{fg:sandwich_grid}, the $z\sim0$ massive main sequence galaxies tend to have a lower $A_{\rm IRX}$ than the $z\sim1$ massive main sequence galaxies, though the two samples largely overlap.
The two samples also have quite similar observed $\beta$.
In terms of $\Delta A$, the $z\sim0$ sample has a median $\Delta A = 7.1$ mag and tends to keep $\Delta A<20$, while the $z\sim1$ sample has a median $\Delta A = 22.3$ mag and scatters up to $\Delta A \approx 100$.
The ratio of the median $\Delta A$ of the two samples suggests that the typical star forming clouds at $z\sim1$ have 3 times the column density of the $z\sim0$ star forming clouds.

From the location of a galaxy on the models grid we determine the total UV attenuation from one face of the gas disc (the sandwich) to the other, i.e. through two slabs of gas and an `even-mix' of gas and stars in the middle, 
\begin{equation}
A_{\rm disc}=2A_{\rm min}+\Delta A \quad.
\end{equation}
The median $A_{\rm disc}$ is 8.5 and 23.7 mag in the $z\sim0$ and $z\sim1$ samples respectively.

The medians of the CO-measured mean $\Sigma_{\rm mol}$ in the $z\sim1$ and $z\sim0$ samples are 546 and 45 M$_{\odot}$~pc$^{-2}$, respectively - a ratio of 12.
The ratio of the median $\Delta A$ or the median $A_{\rm disc}$ are $\sim3$, and this is also the ratio of the gas column densities towards the star forming regions.
Thus, if we attribute the rest of the increase in $\Sigma_{\rm mol}$ to the filling factors, we get that $f_{\rm fill}(z\sim1) / f_{\rm fill}(z\sim0) \approx 4$.

%%%%%%%%%%%%%%%%%%%%%%%%%%%%%%%%%%%%%%%%
\subsection{$N_{\rm H}/A_{\rm V}$}
\label{subsec:NH/AV}
%%%%%%%%%%%%%%%%%%%%%%%%%%%%%%%%%%%%%%%%
The column density of a foreground screen of gas is related to the attenuation of a point source by:
\begin{eqnarray}
\left( \frac{N_{\rm H}}{A_{\rm V}} \right) &=& \left( \frac{A_{\rm disc}}{A_{\rm V}} \right) \left( \frac{N_{\rm H}}{A_{\rm disc}} \right) \nonumber \\
 &=& 0.30^\gamma \left( \frac{N_{\rm H}}{A_{\rm disc}} \right) \quad {\rm cm^{-2}\, mag^{-1}} \, ,
 \label{eq:NH2A1600}
\end{eqnarray}
where $A_{\rm V}$ is the attenuation through the galaxy disc in the optical V band, and $\gamma=-1.18$ (N13) is the UV attenuation law ($A\propto\lambda^\gamma$), that we will assume to extend from 1600~\AA\ up to 5400~\AA.
$A_{\rm disc}$ is the mean attenuation perpendicular and through the galaxy disc which we defined in \S~\ref{sec:IRX-beta_location}.

A molecular gas-mass surface density $\Sigma_{\rm mol}$ can be translated into hydrogen column using:
\begin{equation}
N_{\rm H} = 9.3\times10^{19} \left( \frac{\Sigma_{\rm mol}}{M_{\rm sun} \cdot {\rm pc}^{-2}} \right) \quad {\rm cm^{-2}} \, ,
\label{eq:NH2Sigma_mol}
\end{equation}
where we assumed $\mu_{\rm H}=1.35$ including helium.
Combining Eq.~\ref{eq:NH2A1600} \& \ref{eq:NH2Sigma_mol} we can use $\Sigma_{\rm mol}$ measured from CO, and $A_{\rm disc}$ derived from the $A_{\rm IRX}$--$\beta$ models grid above to calculate $N_{\rm H}/A_{\rm V}$ in the sampled galaxies.
For comparison, the $N_{\rm H}/A_{\rm V}$ ratio in the Milky Way is \citep{Bohlin1978}
\begin{equation}
\left( \frac{N_{\rm H}}{A_{\rm V}} \right)_{\rm MW} = 1.87\times10^{21} \quad {\rm cm^{-2}~mag^{-1}} .
\end{equation}

\begin{figure}
 \centering
 \includegraphics[width=\columnwidth]{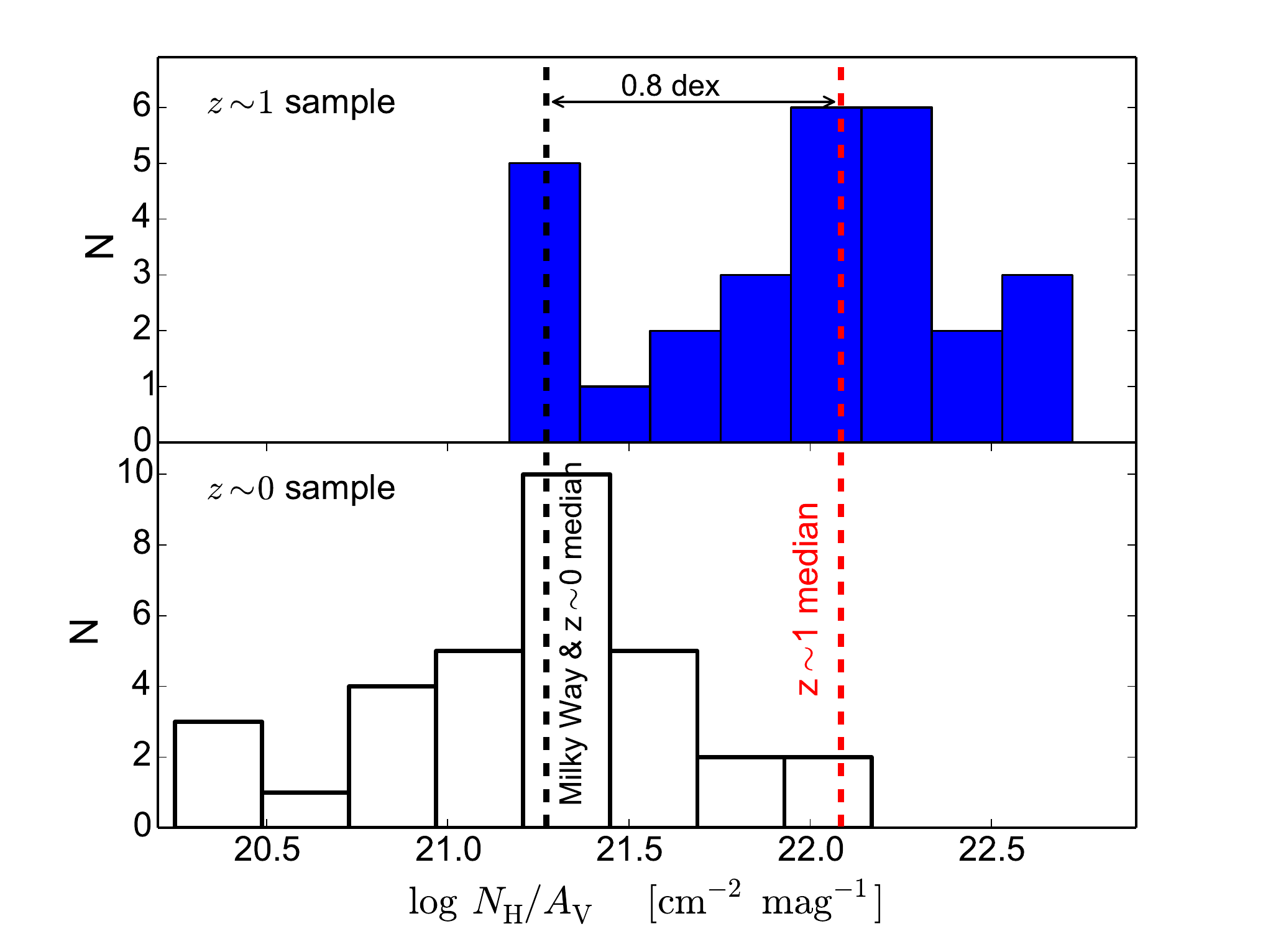}
 \caption{Histograms showing the derived $N_{\rm H}/A_{\rm V}$ in the $z\sim1$ sample ({\it top}) and the $z\sim0$ sample ({\it bottom}).
 The median for the $\sim0$ sample is practically identical to the Milky-Way value.}
 \label{fg:NH2AV_hist}
\end{figure}

Figure~\ref{fg:NH2AV_hist} shows a histogram of the $N_{\rm H}/A_{\rm V}$ values in the $z\sim1$ and $z\sim0$ samples.
While the $z\sim0$ sample shows a rather wide scatter, the median is identical to the Milky-Way value.
The $z\sim1$ sample median ($1.22\times10^{22}$ cm$^{-2}$ mag$^{-1}$) on the other hand is significantly higher than the Milky-Way value by about 0.8~dex.
In other words, the $z\sim1$ galaxies, while having somewhat higher $A_{\rm IRX}$ (Figure~\ref{fg:sandwich_grid}), exhibit low attenuation for their gas columns.

It is important to note that $A_{\rm disc}$ measures the UV attenuation through the galaxy disc at the locations of the star formation, while $N_{\rm H}$ (derived from $\Sigma_{\rm mol}$) is averaged over the area of the galaxy, including regions without any star formation.
When we average $N_{\rm H}$ over the galaxy disc we lower $N_{\rm H}$ by a factor $f_{\rm fill}$ which is the filling factor of the gas.
This suggests that all else being equal, the molecular gas filling factor of the $z\sim1$ galaxies is larger by a typical factor of 6 than in the $z\sim0$ galaxies,
a little higher than the factor 4 we estimated in \S~\ref{sec:IRX-beta_location}.
Given that the `clumpiness' factor (approximately the inverse of the filling factor) is estimated by some authors to be between 5 and 7 \citep{Krumholz2011, Leroy2013}, this means that the filling factor of the $z\sim1$ galaxies is close to 1.
Such high $f_{\rm fill}$ are plausible in galaxies whose molecular gas masses are as large or larger than their stellar masses \citet{Daddi10, Tacconi13, Scoville2016}, often by a significant margin.

In the $z\sim0$ galaxies the effective $N_{\rm H}/A_{\rm V}$ is about the same as Milky-Way, however we must take into account that $f_{\rm fill}<1$ by multiplying $\Sigma_{\rm mol}$ in Eq.~\ref{eq:NH2Sigma_mol} by $f_{\rm fill}^{-1} \approx 6$ and assume that the molecular gas and star formation are co-located.
Thus, after accounting for $f_{\rm fill}$ we get that $N_{\rm H}/A_{\rm V}$ towards the star forming regions is typically higher than Milky-Way by a factor of $\sim6$ in both samples.
Since the mean $\Sigma_{\rm mol}$ increase by factor 12 from $z\sim0$ to $z\sim1$, the increase in $f_{\rm fill}$ suggests that the column densities of the clouds increase by a factor of 2 respectively.

The high effective $N_{\rm H}/A_{\rm V}$ in the high redshift galaxies pose an interesting puzzle.
If we adopt the reasonable assumption that the true single line-of-sight $N_{\rm H}/A_{\rm V}$ value in the $z\sim1$ galaxies is close to Milky-Way value, then by reversing the above equations we get that the column densities towards the star-forming regions are median 6 times (0.8 dex) lower than the mean column density in the entire galaxy.
This is counter intuitive since stars are forming in high density regions and one may expect that the typical column towards the young stars will be higher than the overall mean, not lower.
We shall discuss this further in \S~\ref{sec:discussion}.

%%%% % % % % % % % % % % % % % %
\subsection{The specific attenuation}
% % % % % % % % % % % % % % % % %
The results of the previous section seem to suggest that the gas in high redshift galaxies is less efficient at obscuring the young stars than the gas in low redshift galaxies.
In order to better understand this result we should look at how much obscuration does a unit of gas provides.
For this purpose N13 defined the `specific attenuation' as
\begin{equation}
S_A \equiv \frac{ A_{\rm IRX} }{ M_{\rm mol}/{\rm SFR} } = \frac{ A_{\rm IRX} }{ \tau_{\rm dep} } \, .
\label{eq:SA}
\end{equation}
$S_A$ represents the effective attenuation per gas mass available per young star.
For practical reasons, we use the SFR instead of the number of UV emitting stars, the two being proportional under the assumptions also required for the UV and FIR luminosities conversions to SFR \citep{Kennicutt98}.
The denominator is then equal to the gas depletion time-scale.
$S_A$ is sensitive to the surface density and the geometrical arrangement of the gas and stars.

\begin{figure}
\centering
\includegraphics[width=\columnwidth]{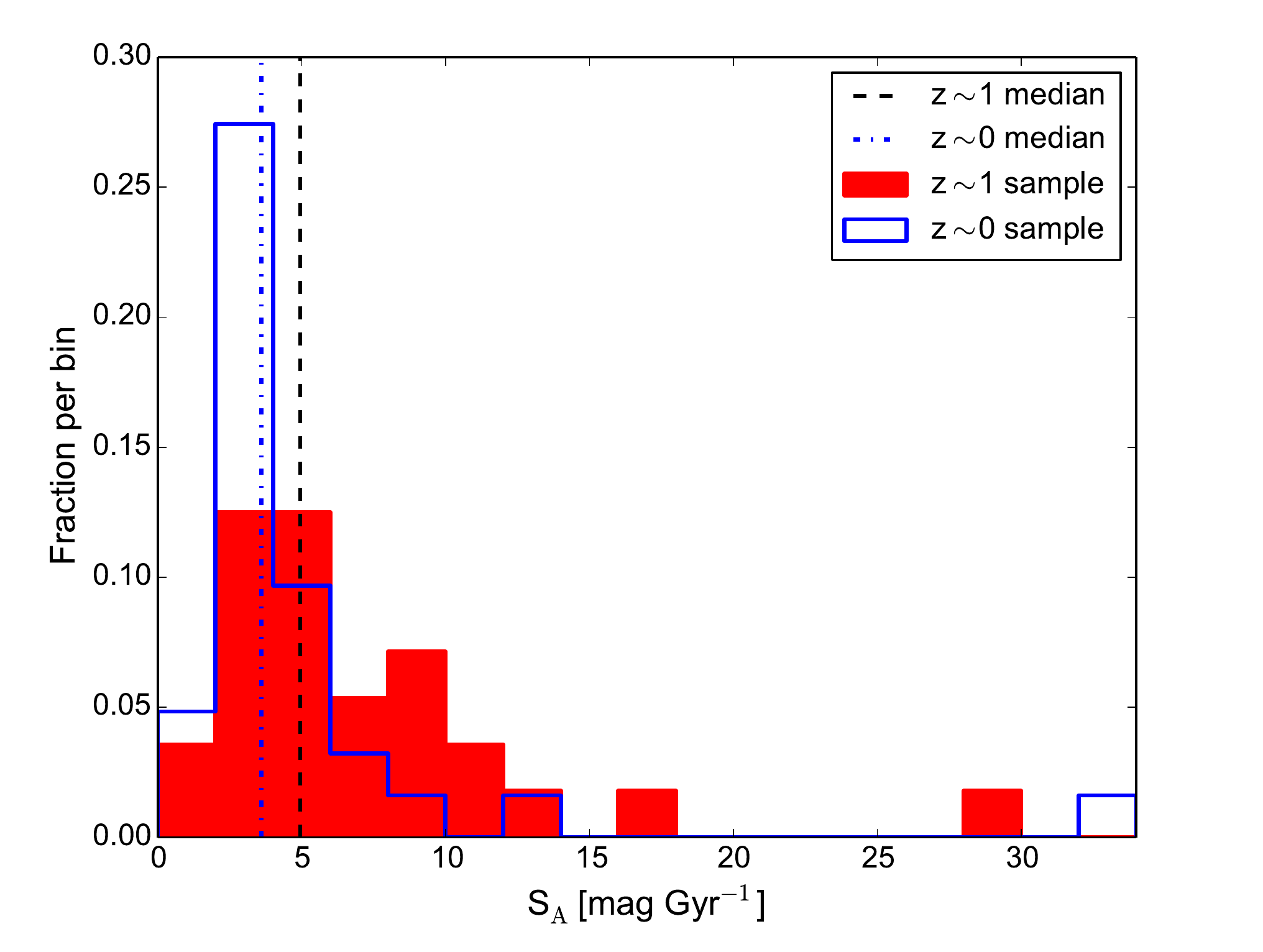}
\caption{Histograms of the specific attenuation in the $z\sim1$ (red) and $z\sim0$ (blue) samples in 2 mag~Gyr$^{-1}$ bins.
		The broken vertical lines indicate the median for each sample. }
\label{fg:SA_histogram}
\end{figure}

In Figure~\ref{fg:SA_histogram} we plot the normalized histograms of the $S_A$ in our two samples.
While the $S_A$ in the $z\sim1$ sample is a bit more widely scattered, the medians of the two distributions are close: 4.9 mag~Gyr$^{-1}$ in the $z\sim1$ sample versus 3.6 mag~Gyr$^{-1}$ in the $z\sim0$ sample.
For reference, $S_A$ in local ultra luminous infrared galaxies (ULIRGs) and $z>1$ sub-millimeter galaxies (SMGs) are an order of magnitude higher (N13).
The median $\tau_{\rm dep}$ of the two samples are also quite close, 0.77~Gyr and 0.96~Gyr respectively.
This means that while the the higher redshift galaxies have significantly more gas, they also have about proportionally more young stars and this gas does not provide less obscuration per star than in the low redshift galaxies - it actually provides slightly more obscuration, and this is reflected in our estimations above that the column densities of the $z\sim1$ star forming clouds are about 2--3 times higher than in the $z\sim0$ sample.

The simplest way of increasing the mean surface densities without increasing $S_A$ is to add similar molecular clouds and star forming regions to the disc in the voids between clouds.
In other words, pack more of the same basic units of star formation per area of the disc.
The similar $S_A$ in the $z\sim0$ and $z\sim1$ samples supports the idea that $f_{\rm fill}$ increase by a median factor 4--6 and cloud column densities increase by factor 3--2 in the latter relative to the former, but otherwise star formation happens in quite similar local environments in both samples.

%%%%%%%%%%%%%%%%%%%%%%%%%%%%%%%%%%
%
\section{Discussion}
\label{sec:discussion}
%
%%%%%%%%%%%%%%%%%%%%%%%%%%%%%%%%%%%

Above we found that in comparison to galaxies in the local universe, the $z\sim1$ galaxies have low UV attenuation for their large gas column densities, which is reflected in $N_{\rm H}/A_{\rm V}$ 6 times the Milky-Way value.
In addition, in the $z\sim1$ galaxies the typical gas columns at the locations of star formation is {\it lower} than the mean gas column of the entire galaxy.
What can be the reason for such a discrepancy between the gas columns directly measured from CO and the columns derived from UV attenuation and slope?
We shall argue that the answer requires significant amount of molecular gas that is spatially removed from the lines-of-sight to the young stars.
But first, let us rule out the following three possibilities that do not require that we break the spatial correlation between young stars and gas density:
\begin{enumerate}
\item Low metallicity
\item $A_{\rm disc}$ is under estimated
\item Measured $A_{\rm IRX}$ is biased low
\end{enumerate}

(i) First possibility considered is that the metallicity is low and thus $N_{\rm H}/A_{\rm V}$ increase.
However, all the galaxies in our sample are rather massive.
According to the stellar masses in \citet{Tacconi13}, the PHIBSS galaxies in our sample have a median $\log M_{\rm star} = 10.8$ solar and a minimum $\log M_{\rm star} = 10.4$ solar.
In such galaxies at high redshifts we expect a metallicity of $Z^\prime \sim 0.8$ solar \citep{Erb2006a}, as measured from the nebular lines ${\rm H}_\alpha/[{\rm N}_{\rm II}]$ ratio.
The $z\sim0$ sample has a similar median $\log M_{\rm star} = 10.5$ and given the lower redshifts are even less likely to have metallicities much below solar.
Based on local studies \citep[e.g.,][]{Draine2007, Leroy2011a, Remy-Ruyer2014} it is usually assumed that at the relevant $Z^\prime$ range the gas-to-dust ratio is about proportional to $Z^{\prime-1}$.
Redshift $z\gtrsim1$ studies tend to agree with this assertion and find that gas mass estimates based on dust emission are consistent with direct CO measurements \citep{Genzel2015, Berta2016}, ruling out unusually high gas-to-dust ratios at $z\sim1$. 
The near solar metallicity in our sample is hardly low enough to explain a factor of 6 increase in $N_{\rm H}/A_{\rm V}$.

(ii) Another possibility is that the $A_{\rm disc}$ is under-estimated.
\citet{Wuyts11b} and \citet{Genzel2013} assumed a pure `even-mix' geometry without an additional foreground screen to convert $A_{\rm IRX}$ in the former, or the attenuation in H$_\alpha$ in the latter, to the total attenuation through the disc.
Such a treatment will indeed increase the derived $A_{\rm disc}$.
This can be seen in Figure~\ref{fg:sandwich_grid} by shifting the galaxies horizontally (constant measured $A_{\rm IRX}$) to the $A_{\rm min}=0$ limit, thus reaching a higher $\Delta A$ and as a result a higher $A_{\rm disc}$ value.
\citet{Wuyts11b} and \citet{Genzel2013} results are still equivalent to a higher than Milky-Way $N_{\rm H}/A_{\rm V}$, by factors of 2--5.
When we do this exercise on our $z\sim1$ sample we find a median  $A_{\rm disc} = 37$, that translates to a local $\Sigma_{\rm mol}=180$ M$_\odot$~pc$^{-2}$, still a factor of 3 lower than the median $\Sigma_{\rm mol}=546$ M$_\odot$~pc$^{-2}$ measured from CO.
In any case, a pure `even-mix' without any foreground component is inconsistent with the UV slope of these galaxies, where the observed $\beta$'s are too high.

(iii) Can it be that the $A_{\rm IRX}$ that we measure is too low?
It is not likely that we are missing FIR flux.
Disc-like galaxies, even high redshift ones are at a low optical depth in FIR wavelengths and thus emit isotropically.
The UV emission on the other hand is much more dependent on geometry.
Little UV is emitted in the plane of the disc relative to the perpendicular direction and we treat the surface densities as if observed with the disc being face-on.
Indeed nearly all the galaxies in our sample are seen to be close to face-on in the {\it Hubble} images \citep[thumbnails availble in][]{Tacconi13, Magnelli12, Daddi10}.
A small number of galaxies are amorphous in their shape or interacting and none are clearly observed edge-on (by our selection).
%If any of the galaxies are observed edge-on it means that we over-estimate their $A_{\rm disc}$ as it would have been measured face-on, and their real $N_{\rm H}/A_{\rm V}$ is possibly even higher.
We do not find any correlation between the location of the galaxy on the $A_{\rm IRX}$--$\beta$ plane
and its morphology or its derived $N_{\rm H}/A_{\rm V}$,
within the $z\sim1$ sample.

UV reflection can increase the UV brightness and lower the derived attenuation.
Without invoking any special geometries, the upper limit would be a factor of 2 higher UV luminosity due to reflected light.
This translates to a decrease of derived $A_{\rm IRX}$ by $\approx 0.75$ mag (Eq.~\ref{eq:A_IRX_definition}).
If we compensate for this and repeat all the calculations, we get a median $N_{\rm H}/A_{\rm V} = 6.9\times10^{21}$ cm$^{-2}$~mag$^{-1}$ for the $z\sim1$ sample, still a factor $\sim$3.5 higher than the Milky-Way.

After ruling out the possibilities above, we have to revisit the assumption that goes back to the KS-relation and in which the densities of the young stars (hence the local SFR density) and gas are spatially correlated, or 'mixed' within any given volume.
In more technical terms, what we implicitly assumed so far is that $f(A_{\rm stars})$ for the stars represents $f(A_{\rm gas})$ in the gas with a similar filling factor.
If we allow the volumetric ratio of SFR and gas density to vary, we can imagine two scenarios that can potentially explain the high $N_{\rm H}/A_{\rm V}$ in the $z\sim1$ galaxies:
\begin{enumerate}
\item A population of quiescent molecular clouds that show little or no star formation.
\item A population of young stars with little to no gas in their immediate environment, but possibly some gas in the foreground.
\end{enumerate}

(i) We can explain the high $N_{\rm H}/A_{\rm V}$ in the $z\sim1$ galaxies as due to a population of quiescent molecular clouds that show little or no star formation.
In our own galaxy we can observe large molecular clouds that show little star formation \citep[e.g.,][]{Maddalena85, Mooney88}.
If the $z\sim1$ galaxies can maintain a large population of young molecular clouds that have not yet started to form stars at a vigorous rate, we will measure a high mean gas surface density in the galaxy, without an increase in the attenuation.
The latter is measured only towards the star-forming regions, and the result is a high $N_{\rm H}/A_{\rm V}$.

There are problems with the quiescent clouds scenario though.
When accounting for the filling factor, we find about similarly high $N_{\rm H}/A_{\rm V}$ in our $z\sim0$ sample.
$\tau_{\rm dep}$ and $S_A$ within the two samples are about the same as well.
So, a similar relative population of quiescent molecular clouds would need to exist in both samples, and the total mass of the quiescent clouds needs to be about 5 times the mass of the star forming clouds.
In the Milky-Way the lifetime of clouds is estimated to be between 20 and 40 Myr, with a period of star formation that is of 10--20 Myr before dispersing the cloud \citep{Elmegreen1991a}.
Therefore the population of quiescent clouds cannot be larger than that of the star forming clouds, and very unlikely larger by a factor of 5 in other low redshift massive spirals.
While we are unable to conduct a census of quiescent and star-forming molecular clouds in high redshift galaxies, we will argue that by analogy to local galaxies and the Milky Way, a mass of quiescent clouds 5 times the mass of star forming clouds is unlikely.

(ii) A different scenario is that dense populations of young stars in star forming regions blow away much of the obscuring gas and reside in local `holes' of low obscuration.
Local low obscuration could take the form of Swiss-cheese like holes the OB stars punched in their GMCs via winds and supernovae, or a cloud-scale volume the young stars cleared and out-lasted their birth clouds that have been dispersed.
Both situations can be observed in the Milky Way.
In this case, 
in order to get a {\it mean} $A_{\rm V}$ that is a factor 6 lower than $A_{\rm V}$ at the locations of the more obscured stars,
the population of young stars residing in holes needs to be $\sim$5 times the population of young stars still obscured by their clouds.

A significant fraction of the unobscured young stars must still be associated with existing clouds.
An assumption of a large population of massive stars that have dispersed their clouds altogether will result in either a filling factor much lower than $f_{\rm fill}=1$, or a too large population of quiescent clouds (see above).
Also, some gas and dust must still exists in the foreground in order to explain the observed $\beta$ values.
Note that the consequence of obscuration holes is not that UV sources are at variable optical depths, as we already account for that with $f(A)$.
The consequence is that $f(A_{\rm stars})$ for the UV stars is not the same as $f(A_{\rm gas})$ for the gas.
The UV attenuation then no longer accurately traces the gas surface density.

\citet{Genzel2013} showed resolved maps of a main-sequence $z\sim1.5$ galaxy in CO and parameters such as SFR density and $A_{\rm V}$ derived from SED fitting (their Figures 2 \& 7).
In these maps the CO peak is offset from the peak in (attenuation corrected) SFR density and $A_{\rm V}$.
Even more significant, the attenuation at various location in the galaxy ($\sim$2~kpc resolution) is consistent with a typical $N_{\rm H}/A_{\rm V}$ 5 times the Milky-Way value.
This would mean that the assumed holes are unresolved at $\sim$2~kpc resolution, which may be expected if one assumes that holes will be no larger than the size of the molecular clouds that formed a young clusters.

If much of the UV is emitted from young stars who blew away their obscuring gas, perhaps this could be better resolved in local galaxies.
\citet{Boissier2007} studied resolved H$_2$ maps of local galaxies.
They find a very poor correlation between the gas surface density (HI, H$_2$ and HI+H$_2$) and the UV attenuation on sub-galactic spatial scales.
\citet{Mao2012} studied spatially resolved $IRX$--$\beta$ maps of local spiral galaxies.
They find that the UV clusters tend to have a slightly lower $L_{\rm IR}/L_{\rm UV}$ (and hence, lower $A_{\rm IRX}$) than the galaxy mean background around them.
These results are consistent with the assumption that UV sources reside in obscuration `holes'. 
\citet{Lee2015} on the other hand did find a correlation between the CO intensity (proportional to the molecular gas surface density) and $A_{\rm V}$ on spatial scales as small as 10 pc in the Milky-Way and the two Magellanic clouds, though with a large scatter.

The main difference between the $z\sim1$ galaxies and the $z\sim0$ galaxies is that the filling factor in the formers is much higher.
In the $z\sim0$ galaxies $f_{\rm fill}<<1$ and so, the mean gas column (surface density) of the entire disc is low.
Blowing away some of the local gas that obscures the young stars still leaves them with a local gas column that is higher or at least comparable to the galaxy mean.
This is in contrast to the case $f_{\rm fill}\approx1$, where blowing away some of the obscuring gas can quickly lower the column to below the galaxy mean.
When we account for $f_{\rm fill}$, we end up with a similar $N_{\rm H}/A_{\rm V}$ excess as in the $z\sim1$ galaxies.
This is most likely a result of similar porous structures to the molecular clouds that surround the young stars, but with generally larger clouds at $z\sim1$ that result in moderately higher $A_{\rm IRX}$, higher $S_A$.

%%%%%%%%%%%%%%%%%%%%%%%%%%%%%%%%%%%%
%
\section{Conclusions} \label{sec:conclusions}
%
%%%%%%%%%%%%%%%%%%%%%%%%%%%%%%%%%%%%

In this work we have studied the relation between the UV attenuation, UV color (i.e., the $A_{\rm IRX}$--$\beta$ diagram), and the molecular gas surface densities in a sample of 28 $z\sim1$ galaxies.
These represent the typical, massive, disc-like star forming galaxies at these redshifts and we compared them to a reference sample of similarly massive $z\sim0$ galaxies.
Both samples are dominated by galaxies near the main SFR--$M_{\rm star}$ relation at their respective redshifts and represent the main mode of star formation in the universe \citep{Rodighiero11}.

Using our larger sample we verified the results of N13 that the $A_{\rm IRX}$--$\beta$ location of disc-like galaxies correlates with their gas depletion time scale $\tau_{\rm dep}$.
We used a `sandwich' model to convert the location of the galaxies on the $A_{\rm IRX}$--$\beta$ plane to the local gas column density at the locations of star formation.
We then compared these columns to the galaxies mean gas column densities as measured from CO observations.

We can summarize our main findings as follows:
\begin{enumerate}
 \item The location of disc galaxies on the $A_{\rm IRX}$--$\beta$ diagram correlates with the gas depletion time-scale of the galaxy.
 This allows us to use $A_{\rm IRX}$--$\beta$ to predict $\tau_{\rm dep}$ (Eq~\ref{eq:tdep_from_SA}) of $z\sim1$ disc galaxies with an accuracy of $\sigma\sim0.16$~dex.
 
 \item The galaxies in the $z\sim1$ and $z\sim0$ samples exhibit very similar obscuration properties in their location on the $A_{\rm IRX}$--$\beta$ diagram and in their specific attenuation.
 This suggests similar obscuration geometries and quite similar basic star-formation units that at $z\sim1$ fill the galaxy with a higher filling factor.

 \item We find very high $N_{\rm H}/A_{\rm V}$ values in the $z\sim1$ sample, a median factor of 6 higher than in $z\sim0$ galaxies and the Milky-Way.
 We estimate the gas filling factor is higher by factor 4--6 in the $z\sim1$ sample relative to the $z\sim0$ sample.
 The typical column density of the star forming clouds increase by a factor 3--2 accordingly.

 \item Given the high gas surface densities in $z\sim1$ galaxies, the relatively low UV attenuation means that the UV sources lie in regions where the line-of-sight gas columns densities are actually lower than the galaxy disc mean column density.
 
 \item We rule out most possibilities for a genuinely high $N_{\rm H}/A_{\rm V}$ and favour a scenario in which the high global $N_{\rm H}/A_{\rm V}$ is due to the UV sources blowing away some of the obscuring gas and residing in local holes in their GMCs.
 
\end{enumerate}

After accounting for the lower filling factor, the local sample resembles the $z\sim1$ sample in the mean $N_{\rm H}/A_{\rm V}$, again indicating that the (porous) geometries of the star forming clouds are quite similar, with the $z\sim1$ galaxies having clouds of higher mean column densities, i.e., larger clouds.

%%%%%%%%%%%%%%%%%%%%%%%%%%%%%%%%%%%%%%%%%%%%%%%%%%
%%%%%%%%%%%%%%%%%%%%%%%%%%%%%%%%%%%%%%%%%%%%%%%%%%
%\acknowledgements
\section*{Acknowledgments}

PACS has been developed by a
consortium of institutes led by MPE
(Germany) and including UVIE (Austria); KUL, CSL, IMEC (Belgium); CEA,
OAMP (France); MPIA (Germany); IFSI, OAP/OAT, OAA/CAISMI, LENS, SISSA
(Italy); IAC (Spain). This development has been supported by the funding
agencies BMVIT (Austria), ESA-PRODEX (Belgium), CEA/CNES (France),
DLR (Germany), ASI (Italy), and CICYT/MCYT (Spain).

This study makes use of data from AEGIS, a multiwavelength sky survey conducted with the Chandra, GALEX, Hubble, Keck, CFHT, MMT, Subaru, Palomar, Spitzer, VLA, and other telescopes and supported in part by the NSF, NASA, and the STFC.

%%%%%%%%%%%%%%%%%%%%%%%%%%%%%%%%%%%%%%%%%%%%
%\bibliographystyle{apj}
\bibliographystyle{mnras}
\bibliography{bibli}
%%%%%%%%%%%%%%%%%%%%%%%%%%%%%%%%%%%%%%%%%%%%%
% latex x bibtex x latex x latex x

%%%%%%%%%%%%%%%%%%%%%%%%%
%  FIGURES   %%
%%%%%%%%%%%%%%%%%%%%%%%%%

%%%%%%%%%%%%%%%%%%%%%%%%%%%%%%%%%%%%%%%%%%%%%%
%
%\appendix
%
%%%%%%%%%%%%%%%%%%%%%%%%%%%%%%%%%%%%%%%%%%%%%%

%%%%%%%%%%%%%%%%%%%%%%%%%%%%%%%%%%%%%%%%%%%%%%%%%%%%%%%%%%%%%%%%%%%%%%%%%%%%%%%%%
%%%%%%%%%%%%%%%%%%%%%%%%%%%%%%%%%%%%%%%%%%%%%%%%%%%%%%%%%%%%%%%%%%%%%%%%%%%%%%%%%
%%%%%%%%%%%%%%%%%%%%%%%%%%%%%%%%%%%%%%%%%%%%%%%%%%%%%%%%%%%%%%%%%%%%%%%%%%%%%%%%%

\end{document}